\documentclass[runningheads]{llncs}

\usepackage[T1]{fontenc}
\usepackage{graphicx}
\usepackage{xcolor}
\usepackage{multicol}
\usepackage{multirow}
\usepackage{url}
\usepackage{makecell}
\usepackage{amsmath}    
\usepackage{amssymb}    
\usepackage{bm}         
\usepackage{listings}
\usepackage{mathtools}

\newif\ifusevspace
\usevspacefalse 

\let\origvspace\vspace
\let\origlinespread\linespread

\renewcommand{\vspace}[1]{\ifusevspace\origvspace{#1}\fi}
\renewcommand{\linespread}[1]{\ifusevspace\origlinespread{#1}\fi}

\begin{document}
\linespread{0.95}



\title{Optimizing Teacher-Student Partitioning for Scalable Knowledge Distillation on HPC Systems }

\titlerunning{Optimizing Teacher-Student Partitioning for Scalable KD}

\author{Adrian P. Dieguez \and
Victor Conchello Vendrell \and
Alex Batlle \and Vinnam Kim \and Jordi Ros-Giralt \and Harris Teague }

\authorrunning{Adrian P. Dieguez et al.}

\institute{Qualcomm AI Research\footnote{Qualcomm AI Research is an initiative of Qualcomm Technologies, Inc.}\\
Correspondence to: \email{aperezdi@qti.qualcomm.com}\\
}

\maketitle

\begin{abstract}
Knowledge Distillation (KD) enables training smaller ``student'' models under the guidance of larger ``teacher'' models, and the widely adopted TRL library implements it. Yet, TRL treats both models symmetrically, missing opportunities to exploit their pronounced asymmetry in memory footprint, and communication requirements. This paper presents an HPC-aware methodology for KD that decouples teacher and student partitioning efficiently. Our approach achieves up to 67\% higher samples-per-second than TRL by avoiding unnecessary teacher-model data structures and selecting the best split strategy. We combine vertical and horizontal partitioning of models, deriving an analytical expression that identifies the existence of inflection points between splitting regimes. These results showed that exploiting teacher--student asymmetry through topology-aware parallelism notably accelerated GKD training on production HPC clusters at our company.

\end{abstract}

\vspace{-3mm}
\section{Introduction}
\vspace{-3mm}

Large Language Models (LLMs) have a huge impact on different scientific and technological applications, accelerating discoveries and productivity. Contemporary LLMs are predominantly autoregressive Transformers \cite{transformers} that generate tokens sequentially. The accuracy of the model is highly correlated to the quality of the training data, and the number of parameters required for training it. Nevertheless, large models have a negative impact on the memory and communication footprint; hence, multiple research efforts focus on getting similar model accuracy by using smaller models, especially on the edge-AI industry.

Knowledge distillation (KD) has become a principal method to compress such models by transferring knowledge from a large teacher to a smaller student. Classic results such as DistilBERT report a 40\% parameter reduction while retaining roughly 97\% of BERT’s accuracy \cite{bert}, motivating widespread adoption of KD. 
Specifically, KD utilizes probabilities produced by the teacher model on the student's loss calculation to accelerate its training by transferring knowledge. 
In some cases, this enables keeping the accuracy of a larger teacher model while reducing its computational cost and size.
For autoregressive LLMs, Generalized Knowledge Distillation (GKD) \cite{gkd} addresses train–inference distribution mismatches, and has become a very common technique. In summary, each KD iteration performs a training step on the student and a forward pass (inference-like) on the teacher to obtain probability distributions for the same batch of data. Training and inference stress systems differently: inference prioritizes low-latency, small-batch execution and is often memory bandwidth bound, whereas training targets large global batches that begin compute-bound but become communication-bound at scale. Training at scale requires efficient gradient synchronization, in addition to parameter and data partitioning, where collective communication (e.g., all-reduce, all-gather) dominates. KD uniquely stresses the system because both the teacher (inference) and the student (training) must be resident and frequently accessed. The teacher runs forward passes at the student's batch sizes, effectively performing inference at far larger batches than typical serving in inference frameworks (e.g. vLLM \cite{vLLM}). In multi-node deployments, this hybrid workload couples teacher-side memory and inference collectives with student-side training collectives, so their joint placement and partitioning across GPUs and nodes determine memory footprint, collective latency, and overall throughput. This coupling is further amplified when the teacher uses model parallelism and the student uses data sharding, or vise versa, as each induces different collective patterns. We observe that the HuggingFace TRL GKD implementation underperforms on multi-node setups for two main reasons. First, the teacher is run under a DeepSpeed \emph{training} engine, which allocates optimizer states irrelevant to inference, wasting GPU memory.   Second, TRL mirrors teacher and student partitioning, leading to suboptimal throughput relative to asymmetric partitions. We analyze these inefficiencies on a \textit{Llama3} workload, removing training-only allocations and broaden sharding and placement choices. Further, by decoupling student--teacher partitioning and developing an analytical cost model, we select inference-optimized partitioning strategies for the teacher while independently choosing the best training strategy for the student, reducing end-to-end time. We apply this methodology to train a production model, achieving up to a $67\%$ speedup relative to TRL implementation. In summary, our contributions are: (i) an analysis isolating multi-node inefficiencies in TRL GKD; (ii) teacher-specific modifications that remove training-state overhead; (iii) a pluggable inference backend for the teacher; (iv) an analytical performance model for demonstrating the existence of inflection points in teacher partitioning; and (v) a large-scale \textit{Llama-3} study demonstrating that asymmetric placement and suggested optimizations deliver up to $67\%$ speed-up.

\vspace{-3mm}
\section{Background: LLM Training and Parallelism}
\vspace{-3mm}

\subsection{LLM Training}

 LLM training optimizes model parameters to estimate a probability distribution over text, typically beginning with large-scale self-supervised \textit{pretraining} followed by task-specific \textit{fine-tuning}. Each training iteration runs a \textit{forward pass} dominated by tensor operations—\textit{multi-head self-attention} whose quadratic cost in sequence length drives memory and compute demand—followed by a \textit{backward pass} to compute gradients and an \textit{optimizer step} (e.g., SGD or Adam) that updates parameters. The forward pass produces a softmax projection (probability distribution over the vocabulary) for the next token, compared to  ground truth to compute the \textit{loss}, which drives gradient and parameter updates. Training proceeds over batches of sequences; larger batches reduce steps per epoch but increase  computational and memory requirements per step due to larger activations and optimizer state. When memory is limited, gradient accumulation simulates larger effective batches by aggregating gradients over multiple forward–backward passes before updating parameters. Knowledge distillation follows the same optimization structure but introduces an additional \textit{teacher} model that guides the training of a smaller model, named as \textit{the student}. In each iteration, the student undergoes a standard forward, backward, and update step, while the teacher performs inference-only forward passes to produce soft targets (probability distributions for the given sequence), syncing at the computation of the \textit{loss}. Therefore, the loss measures how close the student’s predictions are to these teacher distributions. Unlike standard KD, which relies on a fixed dataset of teacher outputs, GKD mixes two signals in the loss calculation: (i) the supervised objective using real labels, and (ii) a distillation objective comparing teacher and student vocabulary distributions on student-generated outputs. This helps correct the mismatch between the data seen during training and the sequences produced at inference time, a limitation of classical distillation.
\subsection{Parallelism strategies and frameworks} 
Modern frameworks such as DeepSpeed \cite{deepspeed} and Megatron \cite{megatron} combine different \textit{parallelism strategies} to balance memory usage, compute efficiency, and communication overhead during training in distributed systems. The literature also refers to this distribution as partitioning or splitting. The two most common strategies are Data Parallelism (DP) and Tensor Parallelism (TP). In DP, each GPU processes a disjoint portion of the input batch while holding a replica of the model; gradients are synchronized via  an \textit{all-reduce} across GPUs \cite{10.1145/3731599.3767699}. DeepSpeed ZeRO (Zero Redundancy Optimizer)~\cite{zero} reduces replication by partitioning model states across GPUs, known as Data Distributed Parallelism (DDP). In ZeRO Stage~1, optimized states are sharded; Stage~2 additionally shards gradients; Stage~3 also shards parameters. Due to this, DDP is also known as \textit{model sharding}. For mixed‑precision training, parameters and gradients are kept in FP16 (2 bytes each), while optimizers such as Adam maintain FP32 master parameters and moment estimates, requiring $K=12$ additional bytes per parameter. With $\Psi$ parameters, then $(2+2+K)\times\Psi$ bytes are distributed across GPUs, excluding activations and temporary buffers. Communication in Stages~1 and~2 consists of a gradient \textit{reduce-scatter} followed by parameter \textit{all-gather}, while Stage~3 requires additional \textit{all-gathers} during the forward pass. In TP, the same batch is replicated across GPUs, but parameters within each layer are partitioned. Transformer models split linear projections in attention and feed‑forward layers along the hidden dimension, distributing matrix multiplications across devices. Each layer therefore requires collective communication to assemble partial activations. Consequently, TP reduces per-GPU parameter memory but introduces frequent, fine‑grained  layer-wise communication that scales poorly across nodes with slower interconnects. DeepSpeed-Inference framework \cite{deepspeedinference} uses TP partitioning for distributing models for inference. DDP thus benefits from large per-GPU batches (to amortize communication), and it is independent of model architecture. In contrast, TP introduces fine-grained layer-wise collective, making performance highly sensitive to bandwidth/latency and hop count.

\vspace{-3mm}
\subsection{Related Work}
\vspace{-3mm}
Google DeepMind introduced Knowledge Distillation in \cite{kd}, and later they released General Knowledge Distillation (GKD) in \cite{gkd}, which is currently implemented in HuggingFace's TRL package \cite{trl}. Beyond TRL, other tools also support KD workflows. Tan et al. present the \emph{GKD-framework}, a general tool that unifies multiple distillation methods and emphasizes memory-constrained execution \cite{tan2023gkd}, which is orthogonal to the system issues we target. Also, torchdistill \cite{matsubara2021torchdistill} provides a modular framework for implementing a wide range of KD methods, but focuses on algorithmic reproducibility rather than the distributed systems. 

For multi-node training, DeepSpeed’s ZeRO family reduces memory redundancy via optimizer/gradient/parameter partitioning \cite{zero}, while Megatron-LM popularizes intra-layer tensor parallelism for large Transformers \cite{megatron}. For inference, DeepSpeed-Inference \cite{deepspeedinference} and vLLM \cite{vLLM} optimize latency/throughput with tensor-parallel and KV-cache–aware designs. Instead, our work analyzes role-specific parallelism (vertical/horizontal for the teacher, vertical for the student), and show that topology-aware, asymmetric partitioning yields consistent throughput gains over symmetric TRL defaults.

\vspace{-3mm}
\section{Testing Setup}
\vspace{-3mm}



The experimental platform consists of 16 nodes with 2TB of RAM and eight NVIDIA H100 GPUs (80GB each), interconnected via NVLink-4, providing up to 900GB/s of intra‑node bandwidth. Each node offers 2×200Gbps storage bandwidth and 4×400Gbps compute bandwidth, and relies on a RoCE‑based inter‑node network.
We consider one representative LLM training workload consisting of an 8B‑parameter, \textit{quantized} LLaMA‑3–based model as student. Training was performed via knowledge distillation from a \textit{floating-point} LLaMA-3‑8B teacher using 1024‑token sequences, using BF16 and the Adam optimizer. To reduce memory footprint, \textit{FlashAttention‑2} was enabled. All experiments were conducted with PyTorch 2.5, DeepSpeed 0.15.4, and TRL 0.13.

\vspace{-3mm}
\section{GKD Analysis and Improvements}
\vspace{-3mm}
\label{gkd-analysis}
 TRL’s GKD implementation trains the student with Distributed Data Parallel (DDP), but also executes the teacher forward pass through the DeepSpeed \emph{training} engine (DDP), as illustrated in Listing~\ref{lst:train} \textit{Line-7}, even when only inference is required for the teacher. In particular, TRL forces a \emph{symmetric} configuration (Listing~\ref{lst:train} \textit{L5-6}): the teacher runs with ZeRO-0 whenever the student uses ZeRO-1/2, and both the teacher and the student use ZeRO-3 when the student is in ZeRO-3. This design simplifies orchestration but prevents intermediate combinations (e.g., a ZeRO-2 student with a partitioned ZeRO‑3 teacher). Since teachers do not maintain gradients or optimizer state in inference, they should not benefit from Stages 1 and 2.
 
\begin{lstlisting}[language=Python, numbers=left, basicstyle=\scriptsize\ttfamily, caption={Teacher's initialization with DeepSpeed configuration in TRL's GKD}, label={lst:train}]
 def _prepare_deepspeed(...):
        deepspeed_plugin = self.accelerator.state.deepspeed_plugin
        config_kwargs = deepcopy(deepspeed_plugin.deepspeed_config)
        ...
        if config_kwargs["zero_optimization"]["stage"] != 3:
            config_kwargs["zero_optimization"]["stage"] = 0
        model, *_ = deepspeed.initialize(model,config_kwargs)
        model.eval()
\end{lstlisting}

Our analysis reveals the central limitation in TRL design: the teacher is executed through the \emph{DeepSpeed training engine}, which allocates optimizer-related buffers even though the teacher performs pure inference. Calling \texttt{model.eval()} and wrapping the forward pass in \texttt{no\_grad()} context prevents updates but does not stop DeepSpeed from allocating optimizer machinery associated with training. As a result, the teacher path incurs avoidable memory overhead that directly restricts the attainable micro-batch size, which reduces training throughput (samples/sec). In our testing setup, this effect becomes evident when the student is configured with ZeRO‑2. In TRL’s default setup, the teacher is forced into ZeRO‑0, leading to out‑of‑memory failures even at modest batch sizes. By contrast, we implement an improvement proposal that modifies this code to enable ZeRO‑2 in the teacher (despite the theoretical absence of teacher gradients or optimizer state needs), allowing these unnecessary tensors to be sharded across GPUs. As shown by the initial two rows of Table~\ref{table-hack}, this modification was sufficient to avoid OOMs and complete training, indicating that optimizer-related buffers are indeed allocated for the teacher, benefiting from Stage 2 distribution, despite these elements being unnecessary for inference.

\begin{table}
\centering
\begin{tabular}{l|c|c|c|c}
\hline
\textbf{Name} & \textbf{Student DS stg} & \textbf{Teacher DS stg} & \textbf{Samples/s} & \textbf{$\mu$-batch size} \\
\hline
Original GKD & 2 & 0 &  OOM &  2\\
\hline
Modified-ADAM & 2 & 2 & 43.4 & 2 \\
Modified-ADAM & 2 & 2 & OOM & 4 \\
Modified-None & 2 & 2 & 53.61 & 4 \\
\hline
\end{tabular}
\caption{Effects of DDP on teacher: (Rows 1-2) Sharding teacher optimizer, compared to the original TRL GKD, reduces memory consumption; (Rows 3-4) decoupling teacher and student configurations so teacher can fully get rid of its optimizer (\textit{None} instead of student's \textit{Adam}) further reduces memory usage, enabling larger batch sizes.}
\label{table-hack}
\end{table}
\vspace{-7mm}

Profiling our testing setup further corroborates this diagnosis. Under the original student–teacher (2–0) configuration on 64~GPUs, we observe a large allocation (29.9~GiB total) originating from TRL’s teacher‑init path and traced the DeepSpeed routines responsible for optimizer setup. With our proposal that modifies the code to enable a (2–2) configuration, the same allocation appears in sharded form (i.e. $\frac{29.9}{64}$-GiB per rank), confirming that these optimizer-associated tensors are distributed with Stage 2. Specifically, tracing down the DeepSpeed initialization routine, we observe this allocation corresponds to teacher‑side fp32 (master) weights\footnote{https://github.com/deepspeedai/DeepSpeed/blob/v0.15.4/deepspeed/runtime/\\bf16\_optimizer.py\#L162C13-L162C101}—unnecessary for teacher inference—whereas with our ZeRO‑2 modification these are partitioned according to the ZeRO strategy.

A second source of overhead arises from the fact that TRL uses a single DeepSpeed configuration file for both teacher and student (Listing~\ref{lst:train} \textit{L2-3}), forcing the teacher to inherit the student’s optimizer settings (currently Adam), which is wasteful that under the current TRL implementation requires an optimizer but does not update its weights. Our second improvement proposal for code modification is to decouple the student-teacher configurations, so we can assign a lightweight optimizer to the teacher or directly overwrite it with a \textit{None} object. The implementation of this proposal, as illustrated in Table~\ref{table-hack}, enables larger micro-batches by avoiding any optimizer-related tensor in the teacher model: while the first improvement proposal described in earlier paragraphs enabled the student to run with Stage 2 at a $\mu$‑batch size of 2—previously impossible in TRL’s original implementation—the additional memory savings from altering the teacher’s optimizer configuration allowed increasing the $\mu$‑batch size to 4. Previous proposals show similar improvement for both \textit{bf16} and \textit{fp16} precision. Additionally, although this study used TRL v0.13 for compatibility, this behavior is part of TRL’s core design and persists as of v0.29.

\vspace{-3mm}
\subsection{Experimental Results}
\vspace{-2mm}

\begin{figure}
    \centering

    \begin{minipage}{0.49\linewidth}
        \centering
        \includegraphics[width=1.1\linewidth]{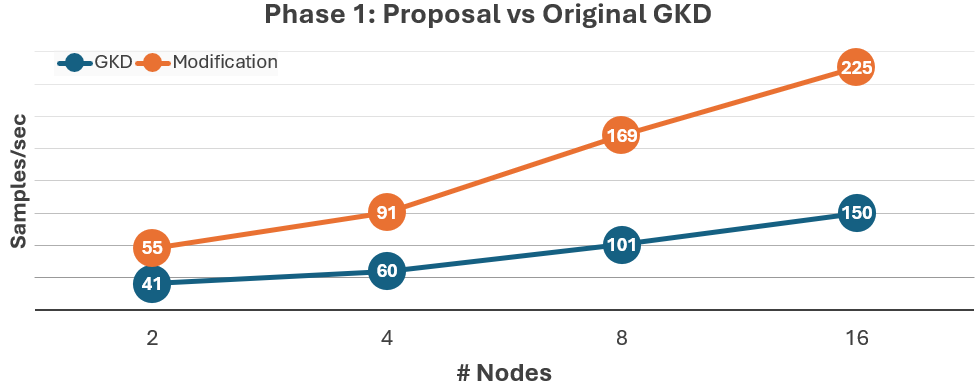}\\
        (a) Throughput
    \end{minipage}
    \hfill
    \begin{minipage}{0.49\linewidth}
        \centering
        \includegraphics[width=1\linewidth]{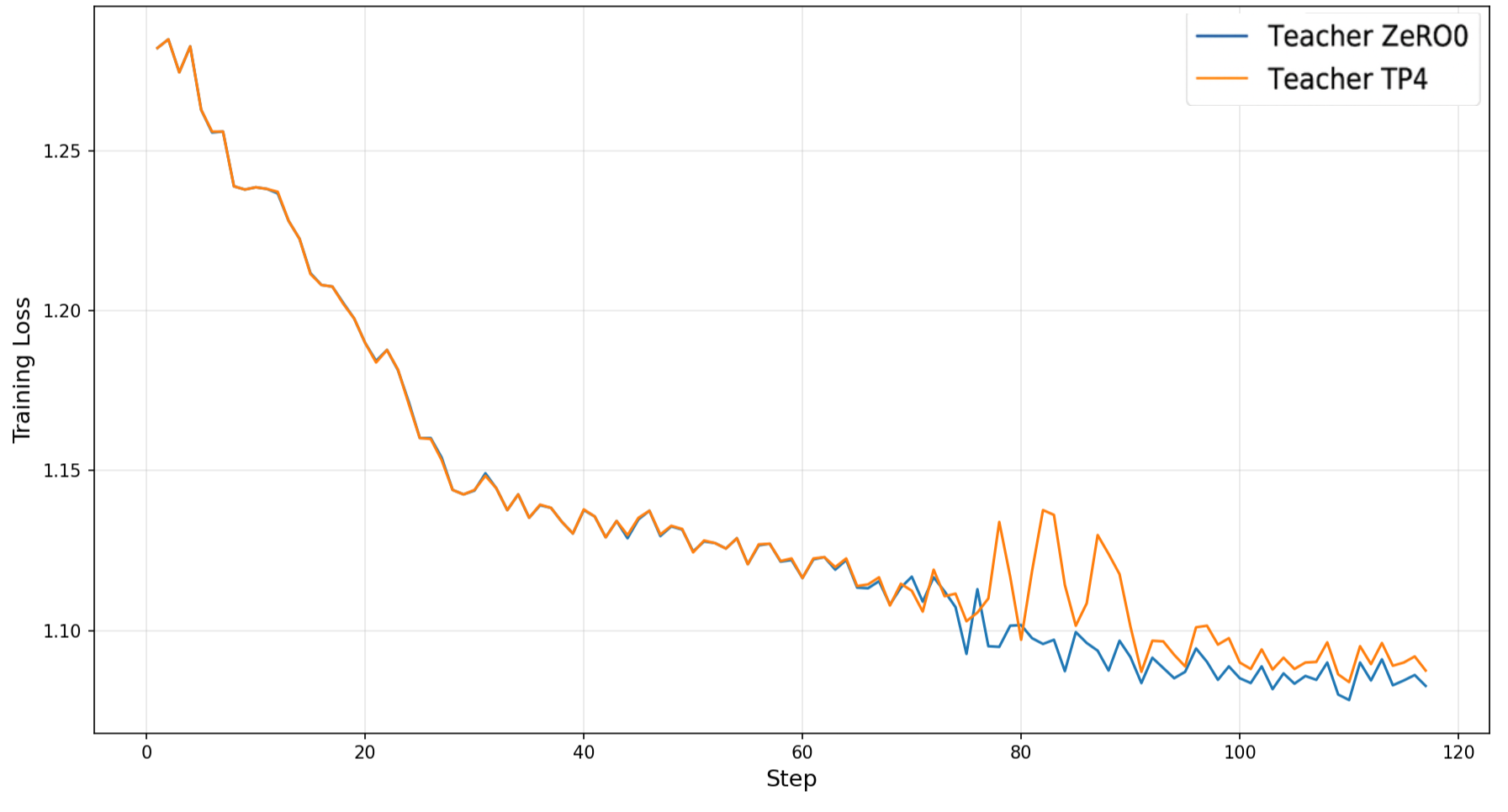}\\
        (b) Accuracy
    \end{minipage}
    \caption{ (a) Training throughput from 2 to 16 nodes comparing original TRL GKD best configuration against our proposal best configuration; (b) Training loss evolution when using ZeRO (DDP) or TP in the teacher, showing no negative effect on accuracy.}
    \label{fig:scaling}
\end{figure}

We implement our improvement proposals and apply them to the \textit{Llama-3} training workflow on our cluster, then conduct an empirical sweep to identify the maximum $\mu$-batch size permitted by each student–teacher DDP pairing—(2–0), (3–0), (2–3), (3–3) in our implementation; and (2–0), (3–0), (3–3) in TRL—using gradient accumulation of 2, \textit{bf16} precision, and scales from 2 (16 GPUs) to 16 nodes (128 GPUs). In the TRL baseline, (3–3) consistently delivers the highest throughput across node counts, enabling $\mu$-batches up to 4× larger than (2–0); this is expected, since the TRL implementation is allocating teacher optimizer states that inflates memory consumption, and partitioning those tensors (Stage 3) offsets its extra communication cost once memory is the dominant constraint. Under our proposal, the same (3–3) configuration admits larger $\mu$-batches than TRL (e.g., on 4 nodes, a maximum of 6 versus 4 $\mu$-batch), yet the best overall performance is typically achieved with (2–0). By removing teacher‑side optimizer allocations entirely, (2–0) avoids parameter partitioning on the teacher and thereby reduces communication collectives. For example, on 8 nodes, (2–0) reaches a $\mu$‑batch of 4 at 168 samples/s, whereas (2–3) reaches 6 at 142 samples/s, and (3–3) reaches 6 at 107 samples/s. In short, once teacher memory overhead is eliminated, communication becomes the limiting factor, and avoiding teacher partitioning yields higher throughput. Finally, as shown in Figure~\ref{fig:scaling} (a), the best-performing configuration of our method, usually (2-0) at the maximum allowed $\mu$-batch, consistently achieves roughly $\approx50\%$ higher samples/s than the best configurations available, often (3-3) at its maximum $\mu$-batch, in the default TRL implementation. Similar conclusions were observed with $fp16$.


\vspace{-3mm}
\section{Effects of Horizontal and Vertical Parallelism for GKD}
\vspace{-3mm}

Section~\ref{gkd-analysis} demonstrated how teacher's forward pass can be accelerated under the existing DeepSpeed DDP-style \emph{training} engine. A natural next question is how the model itself should be partitioned. Based on the DeepSpeed \emph{training} documentation\footnote{https://www.microsoft.com/en-us/research/blog/zero-deepspeed-new-system-optimizations-enable-training-models-with-over-100-billion-parameters/}, parameters in DDP are placed on GPUs in a \emph{vertical} manner: entire layers (or contiguous blocks of layers) reside on a single device, so weights within a layer are not split across GPUs. Conceptually, one can picture the layer laid out left to right and “cut” by vertical lines so that different devices own different layers (\textit{vertical splitting}). Our question is whether a \emph{horizontal} split is preferable for the teacher in GKD: Here, the weights \emph{within} each layer are partitioned across devices so that every GPU holds a slice of every layer. This splitting alternative is known as Tensor Parallelism (TP), and changes the balance of memory-communication in DDP at the cost of more-frequent finer‑grained collectives in each forward pass. Horizontal partitioning may enable larger micro‑batches, although vertical partitioning can reduce per‑layer synchronization. The DeepSpeed \emph{inference} engine (also known as \textit{DeepSpeed-Inference} framework) performs this horizontal partition when distributing models multi-node. While vLLM could be used, accessing logits and integrating with TRL was more challenging, motivating the use of \textit{DeepSpeed‑Inference} for the study. The remainder of this section evaluates these regimes for the teacher under GKD to identify where each partitioning mode is most effective.

\begin{figure}
    \centering

    \begin{minipage}{0.49\linewidth}
        \centering
        \includegraphics[width=1\linewidth]{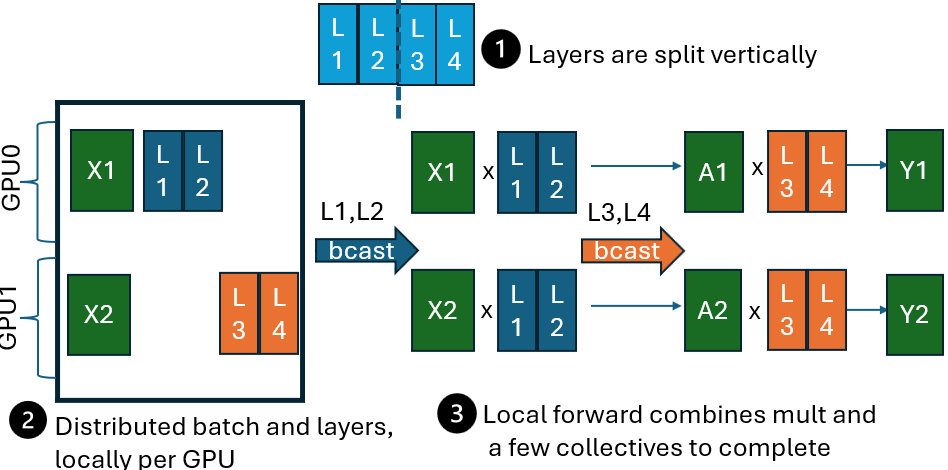}\\
        (a) Vertical split 
    \end{minipage}
    \hfill
    \begin{minipage}{0.49\linewidth}
        \centering
        \includegraphics[width=1.07\linewidth]{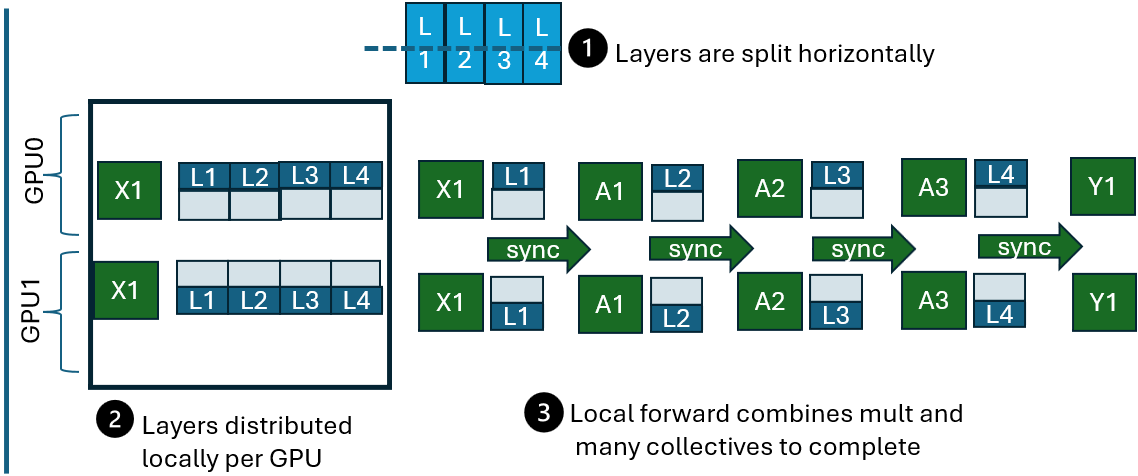}\\
        (b) Horizontal split
    \end{minipage}
    \caption{Partitioning options: (a) partitioning for a forward pass where each GPU holds a disjoint subset of full layers; (b) each GPU stores a portion of each layer, syncing through frequent communication collectives to execute the forward pass.}
    \label{fig:ds_forward_partition}
\end{figure}

Figure~\ref{fig:ds_forward_partition} (a) illustrates the vertical layer partitioning when using DeepSpeed's \emph{training} engine (which uses DDP). In this configuration, full model layers are statically distributed across GPUs to reduce the per-device memory footprint. During the forward pass, every GPU requires access to all model layers in order to process its local data shard under data parallelism. To achieve this, DeepSpeed issues a single collective operation that broadcasts corresponding layer weights. There are different approaches to perform this collective. One option is to do an \texttt{all-gather}, and bring all weights from different GPUs locally into each GPU, but this has a huge memory impact. Another option is to pipeline the communication on several sub-stages to minimize memory footprint, so that each GPU temporarily holds a replica of a subset of layers, which offers the advantage of memory-efficient partitioning and requires only as many collectives as GPUs. In contrast, Figure~\ref{fig:ds_forward_partition} (b) displays the horizontal layer partitioning when using DeepSpeed's \emph{inference} engine for the forward pass, the scheme implemented by Tensor Parallelism (TP). In this setup, every GPU holds a shard of each layer’s weights and processes the \emph{same} batch of input tokens. Because each device computes only a partial output, the intermediate activations must be synchronized across all GPUs before continuing to the next layer. This results in a communication collective (usually an \textit{all-reduce}) after every layer boundary, as illustrated. While TP provides fine-grained model parallelism and enables training very large models by distributing tensors across devices, its major drawback is the high communication frequency: each forward pass incurs one global synchronization per layer, which can become a performance bottleneck on multi-node clusters with limited bandwidth or higher latency. 

Our improved proposal replaces the teacher initialization in TRL (Listing~\ref{lst:train}) with the DeepSpeed \emph{inference} engine (horizontal, TP) rather than the \emph{training} engine (vertical, DDP). Because TP does not shard batches across GPUs while the student trains with DDP, we gather the student’s per‑GPU micro-batches and make the full batch available on each device before invoking the teacher, ensuring both compute the loss on identical inputs while the teacher benefits from inference‑optimized parallelism. To assess whether TP (horizontal) partitioning can outperform DDP (vertical) partitioning in certain GKD regimes, we develop an analytical cost model to identify conditions under which the communication pattern of TP (more frequent but cheaper collectives and typically lower memory per rank) may outweigh the advantages of DDP. This is motivated by the observation that DeepSpeed’s \emph{inference} engine adopts horizontal splitting for inference workloads, whereas the \emph{training} engine relies on vertical splitting for training, and TRL inherits the latter for both teacher and student. Figure \ref{fig:scaling} (b) shows no negative effects on the loss evolution when the teacher uses TP. Our goal is to show that Knowledge Distillation, which combines inference-style teacher computation with training-style student computation, benefits from a hybrid approach rather than a single uniform strategy. We later validate it empirically to demonstrate that there are cases where the hybrid approach (teacher horizontal, student vertical) outperforms the state-of-the-art \textit{static} approach.

\vspace{-3mm}
\subsection{Simple Analytical model of Parallel Training}
\label{analytical}
\vspace{-3mm}
We introduce a deliberately simplified performance model to explain qualitative trends, not to predict exact runtimes. The model isolates how memory footprint, batch size, and communication frequency interact for teacher inference under GKD. Its purpose is to identify regimes where horizontal (TP) partitioning can be preferable to the default vertical (DDP).

\begin{table}[t]
\centering
\begin{tabular}{ll}
\hline
Symbol & Description \\
\hline
$g$ & Total number of GPUs \\
$n$ & Number of nodes \\
$p = g/n$ & Number of GPUs per node \\
$M_g$ & Usable GPU memory for teacher inference, excluding student footprint \\
$M$ & Total model size (parameters only) \\
$a_s$ & Activation memory per sample per layer \\
$b$ & Batch size used by the teacher \\
$t_M$ & Ideal single-GPU compute time per sample \\
$B_{\text{intra}},\, \alpha_{\text{intra}}$ & Intra-node bandwidth and latency \\
$B_{\text{inter}},\, \alpha_{\text{inter}}$ & Inter-node bandwidth and latency \\
$u(b)$ & GPU utilization as a function of per-GPU batch size \\
$\rho$ & Compute/communication overlap factor \\
$A$, $C$ & Latency-, bandwidth-dominated communication terms \\
\hline
\end{tabular}
\caption{Summary of notation for Section \ref{analytical}.}
\label{tab:notation}
\end{table}

Because training performance depends on many interacting factors, any analytical model is necessarily an approximation. Still, this simplified model suffices to reveal that horizontal teacher partitioning can outperform the commonly assumed vertical split in several realistic operating regimes, which we later check empirically. For that, we consider $g$ GPUs arranged into $n$ nodes with $p=g/n$ GPUs per node. Each GPU has available memory $M_g$ for the teacher inference. The model has total size $M$ bytes, $l$ identical layers, and per-sample activation footprint (bytes) $a_s$ (deallocated after use). The ideal single-GPU compute time per sample is $t_M$. Intra- and inter-node links have bandwidth/latency $(B_{\text{intra}},\alpha_{\text{intra}})$ and $(B_{\text{inter}},\alpha_{\text{inter}})$, respectively. GPU utilization is $u(b)\in(0,1]$, depending on per-GPU batch. Compute/communication overlap is modeled via $\rho\in[0,1]$ in the per-batch training time formulation: 
\begin{align}
T_\text{training} \;=\; T_\text{comp} + T_\text{comm} - \rho\cdot\min\{T_\text{comp},T_\text{comm}\}.
\end{align}
All inputs have equal length; memory for matrix--vector ops and allocation overheads are neglected; gradient accumulation is not considered; each layer can be arbitrarily partitioned; and $M_g$ already excludes student-model memory. GPU compute utilization is $u(\cdot)\in(0,1]$ (larger is faster), depending on the number of operations (batch size). To model communication time, we consider all collectives have the same cost for simplicity, which aligns with the hierarchical all-gather implementation. A hierarchical all-gather consists of an intra-node ring, followed by an inter-node ring, and an intra-node scatter (another intra-node ring). If each $i$-GPU contributes $m_i=M/g$ bytes, its cost is
\[
T_{\text{AG}}(m_i)=
2\!\left[(p-1)\alpha_{\text{intra}}+(p-1)\frac{m_i}{B_{\text{intra}}}\right]
+\left[(n-1)\alpha_{\text{inter}}+\frac{n-1}{n}\frac{g\,m_i}{B_{\text{inter}}}\right].
\]
\vspace{-5mm}
Table~\ref{tab:notation} summarizes the notation used throughout the analysis.
\vspace{-5mm}\subsubsection{Vertical split (DDP).}
In the vertical (DDP/ZeRO-style) layout, each GPU owns complete layers, and communication occurs sparsely but with large payloads. In DDP, there is only one single all-gather required to get the layers from other GPUs into local memory and perform the forward operation. However, this would incur an unnecessary memory footprint, and practical implementations pipeline the operation into up to $g-1$ broadcast steps to reduce peak memory usage. Each GPU keeps its $m_i=M/g$ weights and allocates a buffer for handling the reception of other non-local $m_j$ weights. The $\lambda$ parameter, $\lambda\in(0,1]$ (larger values imply larger allocation, and typical values are $1$ or $\frac{1}{(g-1)}$ depending on collective implementation), models this communication buffer. Therefore, the total memory constraint, dominated by weights and activations footprint, determines the batch size:
$M_g> M/g + \lambda (g-1) M/g + a_s b/g \Rightarrow b_{\max}=\lfloor (gM_g-(1+(g-1)\lambda)M)/a_s \rfloor$.

Grouping terms in the $T_{AG}$ expression, we define:
\[
A=2(p-1)\alpha_{\text{intra}}+(n-1)\alpha_{\text{inter}},\qquad
C=2(p-1)\tfrac{1}{gB_{\text{intra}}}+\tfrac{n-1}{n}\tfrac{1}{B_{\text{inter}}}.
\]

where $A$ captures latency-dominated terms and $C$ the bandwidth-dominated terms. Depending on whether the all-gather is pipelined into multiple broadcast stages (normally $\lambda=\frac{1}{(g-1)}$) or executed as a single collective, the resulting per-batch training time expression can be divided by $b$ to obtain the per-sample training time:
\begin{align}
\bar T_{DDP}(b)=
\frac{t_M}{g\,u(b/g)}+\frac{A/\lambda+MC}{b}
-\rho_{DDP}\frac{\min\{T_{DDP\text{comp}},T_{DDP\text{comm}}\}}{b}.
\end{align}
\vspace{-3mm}

\vspace{-5mm}\subsubsection{Horizontal split (TP).}

In the horizontal (TP) layout, each GPU owns a fraction of every layer, leading to frequent but smaller communication events at each layer boundary. The memory constraint is $M_g > M/g + a_s b \implies b_{\max}=\Big\lfloor\frac{M_g-M/g}{a_s}\Big\rfloor$. Each transformer layer performs two communication collectives (after MHA and FFN) with $m_i=a_s b/g$, yielding the resulting per-sample training time time: 
\begin{align}
\bar T_{TP}(b)=
\frac{t_M}{g\,u(b)}+\frac{2l\cdot A}{b}+2l\cdot a_s \cdot C
-\rho_{TP}\,\frac{\min\{T_{TP\text{comp}},T_{TP\text{comm}}\}}{b}.
\end{align}
\vspace{-3mm}

\vspace{-5mm}\subsubsection{Comparison.}

Let $b_{DDP}^*\approx b_{(DDP)}^{\max}$ and $b_{TP}^*\approx b_{(TP)}^{\max}$ be the optimal batches. Assuming utilization saturation near the optimum, $u(b_{DDP}^*/g)\approx u(b_{TP}^*)\approx u_\infty$, and modeling overlap by scaling communication by $(1-\rho_i)$, we obtain
\begin{align*}
\bar T_{DDP}(b_{DDP}^*)\approx \frac{t_M}{g\,u_\infty}
+\frac{(1-\rho_{DDP})(A/\lambda+MC)}{b_{DDP}^*},\\
\quad
\bar T_{TP}(b_{TP}^*)\approx \frac{t_M}{g\,u_\infty}
+\frac{(1-\rho_{TP})\,2l\cdot A}{b_{TP}^*}
+(1-\rho_{TP})\,2l\cdot a_s \cdot C.
\end{align*}
\vspace{-5mm}

Canceling the common compute term, tensor parallelism is faster iff
\vspace{-2mm}
\begin{align}
\boxed{%
\frac{(1-\rho_{TP})\,2l\cdot A}{b_{TP}^*}+(1-\rho_{TP})\,2l\cdot a_s \cdot C
<
\frac{(1-\rho_{DDP})(A/\lambda+MC)}{b_{DDP}^*}
}
\iff\\            
\qquad
\iff
\frac{(1-\rho_{TP})b_{DDP}^*}{(1-\rho_{DDP})b_{TP}^*}
<
\frac{1}{2l}\left(\frac{A/\lambda+MC}{A+b_{TP}^*a_sC}\right)
\label{final-eq}
\end{align}

Solving this inequality to get the inflection points is out of the scope for this paper. However, from the inequality above, we can identify several conditions under which horizontal splitting (TP) may become advantageous. When the model is dominated by weights while activations remain small, the right-hand side grows, making the condition easier to satisfy. A small number of layers has a similar effect, since the right-hand side scales inversely with $l$. Moreover, a high sharding memory overhead (i.e., $\lambda\!\approx\!1$) reduces the maximal batch size $b_{DDP}^*$ ($b_{DDP}^{max}$ depends on memory constraints), enlarging the denominator of the left-hand side and thus penalizing sharded execution. Overlap effects are also important: tensor parallelism benefits when $\rho_{TP}$ is large, whereas $\rho_{DDP}$ is typically smaller, further reducing the left-hand side. Finally, since batch size increases with the number of GPUs, the right-hand side decreases at scale with $b_{TP}^*$, expanding the regime in which tensor parallelism outperforms model sharding.

\vspace{-5mm}
\subsection{Experimental Results}

We empirically found a case where a TP (horizontal) teacher outperforms a DDP-style (vertical) teacher. Specifically, when running the training on a single-node. As shown in Table \ref{tab:single_node_tp_vs_ddp}, with the student at Stage 2 and $\mu BS=3$, the teacher configured with TP surpasses DDP-based teacher variants, which either underperform or run out memory. This is consistent with the intuition that, when traffic is confined to the single-node NVLink fast interconnect, per-layer collectives are not expensive and TP's increased communication is amortized compared to the more expensive communications in DDP. Under our analytical model from Eq. \ref{final-eq}, we find this case satisfies the inequality. The teacher in TP wins over DDP when we assume $\rho_{TP}\!\approx\!\rho_{DDP}$ and $A\!\approx\!0$ (latency negligible relative to bandwidth), yielding:

\[
    \mathclap{
    \frac{4}{3} \sim \boxed{\frac{(1-\rho_{TP})4}{(1-\rho_{DDP})3} < \frac{1}{2\cdot32}\left(\frac{8A+15\text{GB}\cdot C}{A+3 \times 0.0088\text{GB}\cdot C}\right)} = \frac{8A+15\text{GB} \cdot C}{64A+1.6\text{GB} \cdot C} \sim \frac{15\text{GB}}{1.6\text{GB}}.
    }
    \label{eq:tp-faster-condition}
\]
with $1/\lambda = 8$, $l = 32$ and $M = 15\text{GB}$ on \textit{LLama-3-8B}, $b_{DDP} = 4$, $b_{TP} = 3$, $a\_s = 1024 * 4096 * 2 = 0.0088 \text{GB}$. Observe that, as the system scales (more GPUs), $b^*_{TP}$ grows in the denominator of Eq. \ref{final-eq}; even with small $a_s$, this eventually reduces the RHS and the inequality is not satisfied anymore. 

 \begin{table}[t]
\centering
\small
\setlength{\tabcolsep}{6pt}
\begin{tabular}{lcccc}
\textbf{Teacher Mode} & $\boldsymbol{\mu}$\textbf{BS} & \textbf{Student} & \textbf{Teacher} & \textbf{Samples/s} \\
Horizontal & 3 & St~3 & TP & 16.783 \\
Horizontal & 3 & St~2 & \textbf{TP} & \textbf{26.990} \\
Vertical  & 4 & St~2 &  DDP St~2  & OOM \\
Vertical  & 4 & St~2 & DDP St~3        & 21.712 \\
Vertical  & 4 & St~2 & DDP St~3       & 19.696 \\
\end{tabular}
\caption{Single-node (NVLink) results comparing teacher TP to DDP. TP attains the best throughput outperforming DDP-style teacher variants.}
\label{tab:single_node_tp_vs_ddp}
\end{table}

\vspace{-3mm}
\section{Conclusions}
\vspace{-3mm}
This work demonstrates that Generalized Knowledge Distillation can be substantially accelerated by recognizing and exploiting the fundamental asymmetry between teacher and student workloads. TRL’s default GKD implementation executes the teacher through a training-oriented engine, provisioning optimizer states that are unnecessary for the forward pass. Decoupling orchestration, removing training-only allocations on the teacher, and selecting topology-aware partitioning per actor increases throughput by up to 67\% across multi-node runs, enlarge feasible micro-batch sizes, and improve utilization without changing the GKD objective or accuracy. Our results demonstrate consistent $\approx50\%$ throughput improvement from 16- to 128-GPU scenarios.

We formalized the trade-off between vertical (DDP-like) and horizontal (TP-like) partitioning for the teacher and showed the existence of regimes where horizontal splitting, despite its higher synchronization frequency, can outperform vertical splitting due to lower per-rank memory and cheaper per-layer collectives. Importantly, the balance between these regimes is sensitive to interconnect. High‑bandwidth, low‑latency fabrics (e.g., NVLink) favor tensor parallelism by amortizing frequent collectives, whereas  higher‑latency interconnects reduce this advantage and shift the balance toward vertical partitioning. Our analytical model makes this dependence explicit, enabling topology‑aware reasoning rather than assuming a single optimal strategy.

A hybrid strategy (horizontal for the teacher, vertical for the student) emerges as a robust alternative. Empirical results on production \textit{Llama‑3} workloads confirm these predictions and indicate that removing teacher-side optimizer state shifts the bottleneck from memory to communication, after which finding the best teacher partitioning yields the best end-to-end throughput. Finally, the proposed changes are deliberately non‑intrusive. As a result, the approach integrates cleanly into existing TRL workflows with modest orchestration‑level changes, making hybrid parallelism for GKD both practical and immediately deployable.

Future work includes solving and integrating the cost model into autotuning, extending to other types of parallelism (sequence/pipeline), test with long-context, larger-teacher models, and fuse some teacher/student operations.

\begin{credits}
\subsubsection{\ackname} 
Language and grammar improvements were supported by \textit{Microsoft Copilot}. All content, analysis, and conclusions are solely those of the authors.

\end{credits}

 \bibliographystyle{splncs04}
 \bibliography{mybibliography}

@misc{deepspeed,
  author       = {Microsoft},
  title        = {DeepSpeed: Accelerating Deep Learning Training and Inference},
  year         = {2024},
  howpublished = {\url{https://github.com/microsoft/DeepSpeed}},
  note         = {Accessed: 2025-08-06}
}

@inproceedings{transformers,
  title     = {Attention Is All You Need},
  author    = {Vaswani, Ashish and Shazeer, Noam and Parmar, Niki and
               Uszkoreit, Jakob and Jones, Llion and Gomez, Aidan N.
               and Kaiser, {\L}ukasz and Polosukhin, Illia},
  booktitle = {Advances in Neural Information Processing Systems (NeurIPS)},
  year      = {2017}
}

@inproceedings{bert,
  title     = {DistilBERT, a distilled version of BERT: smaller, faster, cheaper and lighter},
  author    = {Sanh, Victor and Debut, Lysandre and Chaumond, Julien and Wolf, Thomas},
  booktitle = {Proceedings of the 5th Workshop on Energy Efficient Machine Learning and Cognitive Computing (EMC$^2$) at NeurIPS},
  year      = {2019},
}

@inproceedings{gkd,
  title     = {On-Policy Distillation of Language Models: Learning from Self-Generated Mistakes},
  author    = {Agarwal, Rishabh and Vieillard, Nino and Zhou, Yongchao and
               Stanczyk, Piotr and Ramos, Sabela and Geist, Matthieu and Bachem, Olivier},
  booktitle = {International Conference on Learning Representations (ICLR)},
  year      = {2024}
}

@article{kd,
  title   = {Distilling the Knowledge in a Neural Network},
  author  = {Hinton, Geoffrey and Vinyals, Oriol and Dean, Jeff},
  journal = {arXiv},
  year    = {2015},
  url     = {https://arxiv.org/abs/1503.02531}
}

@misc{zero,
author = {Rajbhandari, Samyam and Rasley, Jeff and Ruwase, Olatunji and He, Yuxiong},
title = {ZeRO: Memory Optimizations Toward Training Trillion Parameter Models},
howpublished = {ArXiv},
year = {2020},
month = {May}
}

@article{deepspeedinference,
  title   = {DeepSpeed Inference: Enabling Efficient Inference of Transformer Models at Unprecedented Scale},
  author  = {Aminabadi, Reza Yazdani and Rajbhandari, Samyam and Zhang, Minjia and
             Awan, Ammar Ahmad and Li, Cheng and Li, Du and Zheng, Elton and
             Rasley, Jeff and Smith, Shaden and Ruwase, Olatunji and He, Yuxiong},
  journal = {arXiv preprint arXiv:2207.00032},
  year    = {2022},
  url     = {https://arxiv.org/abs/2207.00032}
}

@misc{trl,
  title        = {TRL — Transformer Reinforcement Learning},
  author       = {{Hugging Face}},
  howpublished = {\url{https://huggingface.co/docs/trl/index}},
  year         = {2026},
  note         = {Accessed: Mar 10, 2026}
}

@misc{megatron,
      title={Megatron-LM: Training Multi-Billion Parameter Language Models Using Model Parallelism}, 
      author={Mohammad Shoeybi and Mostofa Patwary and Raul Puri and Patrick LeGresley and Jared Casper and Bryan Catanzaro},
      year={2020},
      eprint={1909.08053},
      archivePrefix={arXiv},
      primaryClass={cs.CL},
      url={https://arxiv.org/abs/1909.08053}, 
}

@article{tan2023gkd,
  title   = {GKD: A General Knowledge Distillation Framework for Large-scale Pre-trained Language Model},
  author  = {Tan, Shicheng and Tam, Weng Lam and Wang, Yuanchun and Gong, Wenwen and
             Yang, Yang and Tang, Hongyin and He, Keqing and Liu, Jiahao and
             Wang, Jingang and Zhao, Shu and Zhang, Peng and Tang, Jie},
  journal = {arXiv preprint arXiv:2306.06629},
  year    = {2023},
  url     = {https://arxiv.org/abs/2306.06629}
}

@inproceedings{matsubara2021torchdistill,
  title     = {torchdistill: A Modular, Configuration-Driven Framework for Knowledge Distillation},
  author    = {Matsubara, Yoshitomo},
  booktitle = {Proceedings of the International Workshop on Reproducible Research in Pattern Recognition (RRPR)},
  series    = {Lecture Notes in Computer Science},
  volume    = {12636},
  pages     = {24--44},
  year      = {2021},
  publisher = {Springer}
}

@inproceedings{vLLM,
author = {Kwon, Woosuk and Li, Zhuohan and Zhuang, Siyuan and Sheng, Ying and Zheng, Lianmin and Yu, Cody Hao and Gonzalez, Joseph and Zhang, Hao and Stoica, Ion},
title = {Efficient Memory Management for Large Language Model Serving with PagedAttention},
year = {2023},
publisher = {ACM},
booktitle = {Proceedings of the 29th Symposium on Operating Systems Principles},
pages = {611–626},
numpages = {16},
series = {SOSP '23}
}

@inproceedings{10.1145/3731599.3767699,
author = {P\'{e}rez Di\'{e}guez, Adri\'{a}n and Batlle Casellas, \`{A}lex and Torres, Aleix and Teague, Harris and Ros, Jordi},
title = {Pretraining LLMs at Scale: Tuning Strategies and Performance Portability.},
year = {2025},
publisher = {ACM},
booktitle = {Proceedings of the SC '25 Workshops of the International Conference for High Performance Computing, Networking, Storage and Analysis},
pages = {1512–1523},
numpages = {12}
}

\end{document}